# A New Approach in Position-Based Routing Protocol using Learning Automata for VANETs in City Scenario

Fatemeh Teymoori[1], Hamid Nabizadeh[1] and Farzaneh Teymoori[2]

[1]Shahid Beheshti University, G. C., Tehran, Iran
`{f.teymor, ha.nabizadeh}@mail.sbu.ac.ir`

[2]Science and Research Branch of Islamic Azad University, Tehran, Iran
`farzanehteymoori@yahoo.com`

## ABSTRACT

*One of the main issues in Vehicular Ad-hoc NETworks (VANETs) is providing a reliable and efficient routing in urban scenarios with regard to the high vehicle mobility and presence of radio obstacle. In this paper, we propose a Position-Based routing protocol using Learning Automata (PBLA). In addition, PBLA uses the traffic information for enhancing learning. As we know, a main characteristic of learning is increasing performance over time. We exploit this characteristic to decreasing use of traffic information. Initially, PBLA make effort to finding best and shortest path to mobile destination using traffic information.*

## 1. INTRODUCTION

Vehicular Ad-hoc Networks (VANETs) is a subclass of Mobile Ad-hoc Networks (MANETs); Vehicles on the roads use wireless technology to communicate each other without any pre-deployed infrastructure. More recently, various applications have appeared in the VANETs. The applications have been classified into two categories: 1. safety applications, which allow the passengers or drivers to share contents such as road obstacles, traffic flows and accidents that have occurred, 2. entertainment applications, which allow vehicles to share multimedia or local information such as MP3 music, videos, sale advertisement or virtual tours of hotel rooms [12]. One of the main issues in VANETs is providing a reliable and efficient routing in urban scenarios with regard to the challenges (i.e., high vehicle mobility and presence of radio obstacle) [1-3].

Generally, the ad hoc network routing protocols are divided in three category: unicast, multicast and broadcast. Unicasts are divided in two parts: topology-based and position-based. Topology-based routing protocols use links information that exist in the network to perform packet forwarding. However, in position-based routing protocols, each node needs only know its neighbors' positions. After using proposed unicast protocols of MANETs in VANETs, it is obvious that these protocols do not work properly in VANETs and they are weak. In addition, most of the protocols that exist in MANETs are topology-based and their major problem is the instability of routes that are caused by link breakages. In topology-based protocols, link breakages occur repeatedly, consequently, the packet loss rate and the overhead of routing increase. Therefore, between topology-based and position-based routing protocols, the last one is more efficient for data delivery in high mobility conditions, such VANETs.

In position-based routing, each node is aware of the positions of its direct neighbors by periodically sending out beacon messages that indicate the current position of the node. In addition, with the aim of sending a packet to a destination node, the sender requires information on the current geographic position of the destination node. This information is gained via a so-called location service [5].

In This paper, a position-based routing protocol in urban scenario is proposed that uses the learning algorithm [11] for decreasing the communication overhead and the number of hops. Furthermore, this protocol uses the traffic information for enhancing learning. This Improving learning reduces the communication overhead that is generated by the traffic information.

In order to gain traffic information, we investigated the movement patterns of the vehicles in the streets in different times of the day and generated several databases for traffic density in different times and areas. For instance, the traffic density is low in rural areas and during night hours but very high in the urban area and during rush hours of the day. Then routing decision with high reliability will be made efficiently.

The rest of this paper is organized as follows: Section 2 describes related work in the field of unicast routing in VANETs. Section 3 reviews the learning automata. Section 4 describes the proposed position-based routing protocol. Simulation results are shown in Section 5. Finally, section 6 concludes the paper and gives directions for future works.

## 2. RELATED WORK

AODV[4] is a simple sample of topology-based. In [4], the authors presented a route discovery phase that the route request packets flood to the network for searching the route. High node mobility leads to disrupted network and the overhead significantly increase due to repairs broken routes.

A well-known position-based routing algorithm is greedy. In this algorithm, the selected next hop node in comparison with the current node is closer to the destination. Greedy does not perform well in urban scenarios because of the radio obstacles, despite the fact that this algorithm has a good performance in creating stable routes in the highways. Some papers have tried to improve position-based routings by using the traffic information [5] [8-10], but the problem is that the traffic information increases the communication overhead.

GPSR [6] is the best-known greedy protocol for ad hoc networks. GPSR uses two methods for forwarding packets to destination: greedy forwarding and perimeter forwarding. Greedy forwarding is used whatever possible and perimeter forwarding is used when greedy failed. As well as, in greedy forwarding, packets are forwarded to nodes that are closer to the destination in Euclidean distance. GPSR uses the perimeter forwarding when there is not a greedy path in some regions of the network. In perimeter mode, a packet traverses successively closer faces of a planar sub-graph of the full radio network connectivity graph, until it reaches a node that is closer to the destination, where greedy forwarding is resumed. The disadvantage of GPSR is increasing the possibility of getting a local maximum and link breakage because of two problems of VANETs. As mentioned, the high mobility of vehicles and specific topological structure of a city [3].

To deal with these problems, a position-based geographic source routing protocol (GSR) was proposed [5]. GSR uses Dijkstra algorithm to calculate shortest path consist of sequence intersections which packet has to traverse.

Forwarding packets is based on greedy forwarding strategy between two successive intersections. The main disadvantage of this scheme is that the shortest path including intersections does not

International Journal of Ambient Systems and Applications (IJASA) Vol.1, No.2, June 2013

mean the best path since hop count is not proper to be taken as a performance metric in high dynamic networks.

GPCR (greedy perimeter coordinator routing) [7] is a position-based routing for urban environment.In highly dynamic environment such as VANETs, GPCR protocol acts well in two scenarios, city, and highway. Packets in GPCR traverse intersections by a restricted greedy forwarding procedure. GPCR proposed a repair strategy based on the topology of streets and intersections for adjusts the routing path.

The main contribution of the scheme is approaches of how to detect vehicles at the intersections without digital map. However, this protocol may lead to redundant hops in city environments because of using right hand rule.

VADD [8] protocol adopted the idea of carry-and-forward for data delivery from a moving vehicle to a static destination. The most important issue is selecting a forwarding path with the smallest packet delivery delay. VADD protocol attempts to keep the low data transmission delay by forwarding packets through wireless channel. In VADD, when a packet needs to carried through roads, the road with higher speed is chosen. VADD assigns cost to edges between each two intersections by proposing delay model to estimate data delivery delay in different roads. In [8] assumes each vehicle is equipped with digital map and traffic statistics such as traffic density and vehicle speed on roads at different times of the day. According to the information, VADD protocol proposed a delay model to assign cost to each edge. With these cost, VADD computes the shortest path from the source to the destination by a naive optimal forwarding path selection algorithm. Disadvantage of VADD is that cannot freely select the outgoing road to forward the packet at each intersection.

The Road-Based using Vehicular Traffic (RBVT) routing [9] leverages real-time vehicular traffic information to create road-based paths. These paths are consisted of successions of road intersections that are found by the flooded route discovery process. According to recorded intersections in the source routing header, geographical forwarding transfers packets between intersections on the path. This protocol increased overhead because of using real-time vehicular traffic information.

## 3. REVIEW OF LEARNING AUTOMATA

In the following, we present a brief review of learning automata.A learning automaton (LA) [11] is an adaptive decision-making unit that improves its performance by learning how to choose the optimal action from a finite set of allowed actions through repeated interactions with a random environment. Random environment Inputs called actions and reinforcement signals are responds the actions to the environments. The action probability vector is brought up to date based on the received feedback of environment. Figure 1 demonstrates correlation between the learning automaton and its random environment.

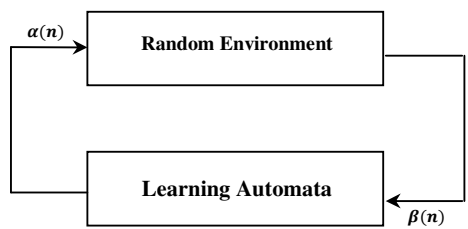

Figure 1. The correlation between the learning automaton and its random environments

Variable structure learning automata is a type of learning automata. Variable structure learning automata are defined by a quaternary $\{\alpha, \beta, L, p\}$ , where $\beta = \{\beta_1, \beta_2, \ldots, \beta_r\}$ is the set of inputs,

International Journal of Ambient Systems and Applications (IJASA) Vol.1, No.2, June 2013

$\alpha = \{\alpha_1, \alpha_2, \ldots, \alpha_r\}$ is the set of actions, and L is learning algorithm. L is a Recursiveequation to modifying action probability vector. Action probability vector is shown by $p = \{p_1, p_2, \ldots, p_r\}$. Equations (1) and (2) show a linear learning algorithm. Let $\alpha_i$ be the action chosen by the automata.

When the selected action is rewarded by the environment (i.e. $\beta(n) = 0$):

$$p_i(n + 1) = p_i(n) + a \cdot (1 - p_i(n)) \tag{1}$$

$$p_j(n + 1) = p_j(n) - a \cdot p_j(n) \quad \forall jj \neq i$$

And, when the selectedaction is penalized by the environment (i.e. $\beta(n) = 1$):

$$p_i(n + 1) = (1 - b) \cdot p_i(n) \tag{2}$$

$$p_j(n + 1) = \frac{b}{r - 1} + (1 - b) \cdot p_j(n) \quad \forall jj \neq i$$

## 4. PROPOSED ROUTING APPROACH

As mentioned before, by investigating traffic behaviors in several days and several times during a day, we concluded that traffic behaviors in street usually repeat unless exceptions happen like the accidents that block the streets.

Therefore, each vehicle can possess several databases for their decision-makings. For instance, one database for rural area and during night, one for urban area and during day, or the other one for exception cases etc. In PBLA we have two phase, Learning and routing phase. When a source node want to send packet, it find the best path to destination using Dijkstra's algorithm and adjacent matrix, which this matrix is built in the learning phase. In the following, we describe two phases of our routing protocols.

### 4.1. Learning Phase

In PBLA, each vehicle can acquire the number of available vehicle in each street by location service. In addition, we consider the street map as a planar graph that is pre-loaded in the vehicles as a text file (intersections as vertexes and streets as edges of this planar graph).

Each vehicle is a Learning Automata (LA) which selects actions by considering input($\beta$) and according to the inputs rewards or penalize them.

Every street will be chosen as an action. The input ($\beta$)for street $S_1$iscalculated by (3).

$$\beta = 0 \; if \; N \geq min \tag{3}$$

$$\beta = 1 \; if \; N < min$$

In this equation,$N$is number of vehicles in $S_1$, which is provided by location service.*min* is the minimum number of the required vehicles for sending through wireless in the specific street in worst case. For example, suppose that there is a street with 200 *m* length. If transmission range is equal 100 *m*, then *min* will be four. When $\beta = 0$ by using (1), the chosen action will be rewarded, and when $\beta = 1$by using (2), the chosen action will be penalized.

Note that$p_i(0) = \frac{1}{number of street}$. Each vehicle does this for all streets.

As a result, each vehicle has a history of probability in form of a vector and it can conclude that how is traffic behavior in that chosen street. Then vehiclescalculate cost of each street base on last probability of that street by (4).

$$Cost = number\ of\ street - \frac{p_i(n+1)}{p_i(0)} \qquad (4)$$

Whatever the amount of probability is greater, then smaller cost will be assigned to each edge. Therefore, each vehicle builds the adjacent matrix for streets by these costs.

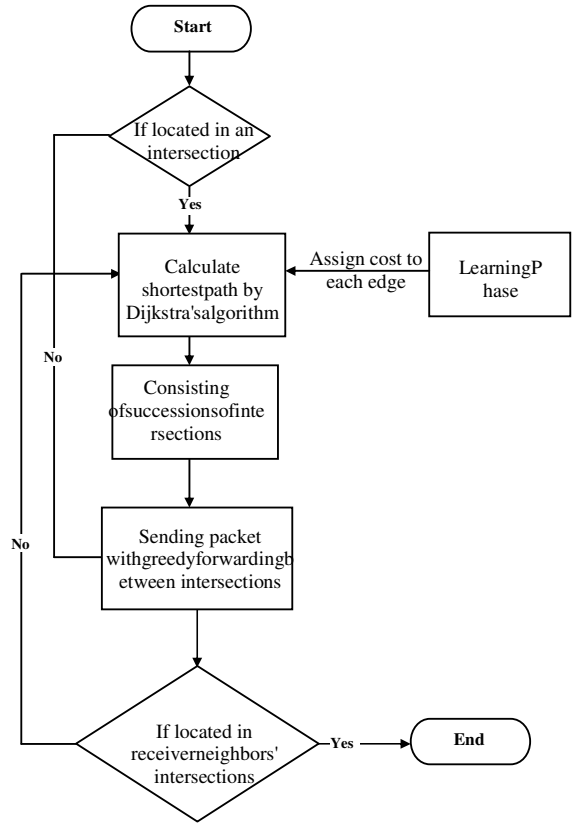

Figure 2. Decision making flowchart

## 4.2. Routing Phase

After we got the probability of each action, inthe routing phase as illustrated in Figure 2, adjacent matrix of graph that intersections are its vertexes and streets are its edges will be updated with the generated cost in the learning phase.

## 5. PERFORMANCE EVALUATION

In this section, we evaluate the performance of our scheme. For illustration of our scheme performance, we compare it with the GPSR [6] and GPCR [7]. For the sake of evaluation, we run simulations on a discrete event simulator, OMNET++ version 4.2.2 [13].

The experiment is based on a 2 × 2.5 km rectangle street area, which presents a grid layout. The scenario consists of 150 vehicles (nodes) in 27 streets and 18 intersections as shown in Figure 3.

49

The street layout is generated by SUMO version 0.15.0 [14]. In addition, we use SUMO for generating random traffic on the streets and intersections. For connecting OMNET to SUMO, we use veins framework [15].

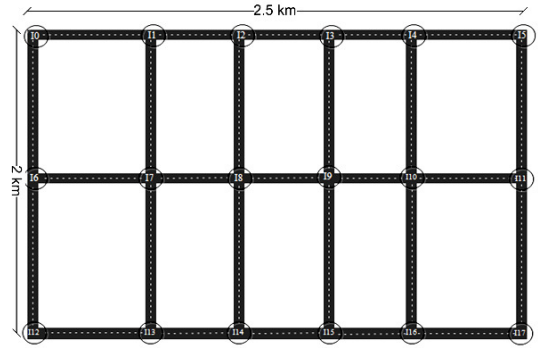

Figure 3. Simulation area

In all of simulations, we considered that transmission range for every vehicle equals to 500 meters. Transmission delay set 18Mbps. MAC layer protocol follows IEEE 802.11p. All experiment parameters are shown in Table 1.

Table 1. Simulation setup

| Parameters | Values |
| --- | --- |
| MAC Protocol | IEEE 802.11p |
| Transmission range | 500m |
| Bitrate | 18Mbps |
| Map | Grid 2×5(2km×2.5km) |
| Number of streets | 27 |
| Number of intersections | 18 |
| Number of vehicle | 150 |
| Packet size | 512 Byte |
| Simulation time | 600s |

For each simulation run, we randomly select ten source-destination pairs. Generating and sending packets is done with uniform distribution by the source nodes. We measured the packet delivery rate as shown in Figure 4 and number of hops as shown in Figure 5 versus the distance between the two source-destination pairs.

In Figure 4 and Figure 5 PBLA compared with GPCR [7] and GPSR [6]. As mentioned before, GPSR has low reliability in urban scenarios due to presence of radio obstacles. For improving GPSR protocol, GPCR forwards packets in greedy mode between each two intersections. Then, whenpackets arrive at intersections, they are forwarded to a node called *coordinator*rather than forwarded across a junction. Therefore, reliability in GPCRhas increased compared to GPSR. As shown in Figure 4, PBLA in comparison with GPSR has a high reliability and compared to GPCR has almost same reliability. This similarity is due to packets in PBLA are forwarded through streets and intersections almost like GPCR.

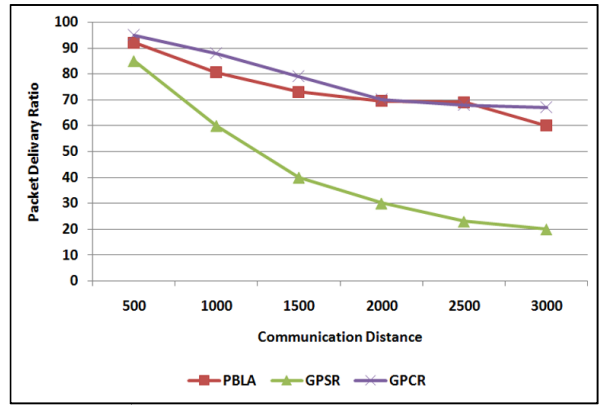

Figure 4. Packet delivery ratio vs. Communication distance

As mentioned in [7],performance improvement in GPCR comes at the cost of a higher average number of hops compared to GPSR. This increase of hop counts is mainly caused by those packets that could not be delivered at all by GPSR and thus did not impact the hop count.

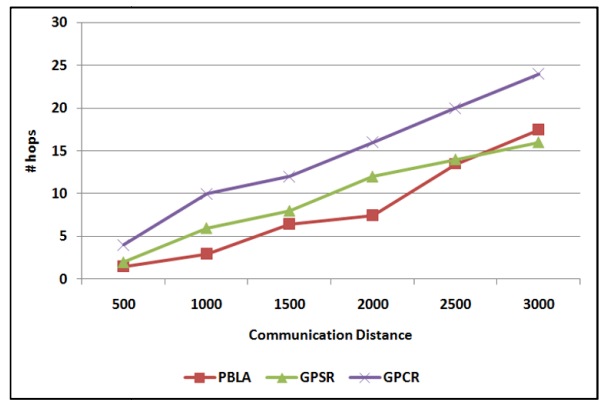

Figure 5. Number of hops vs. Communication distance

As shown in Figure 5,PBLA has minimum average hop count in comparison with two other protocols. PBLAis different from GPCR in the way choosing next intersection for forwarding packet. GPCR find shortest path in terms of distance to next intersection to the destination. However, PBLA using probabilities that gain by learning automata algorithm attempts to find an optimal path to the destination. Path with high possibility forwarding through wireless channel is chosen as an optimal path by learning automata. Note that the optimal path is not necessarily shortest path in terms of distance; nevertheless, the optimal path is a route that packets is sent in least time through channel.

Accordingly, as can be seen in Figure 5, using optimal path that gain by PBLA caused to decreasing in number of hops.

## 6. CONCLUSION AND FUTURE WORKS

In this paper, we presented a position-based routing protocol using learning automata for vehicular ad hoc networks in city scenarios. City scenario and high mobility of vehicular nodes are challenges of VANETs.For covering these challenges, PBLA forwards packets through streets and intersections.Packetsare forwarded in greedy mode between each two intersections until arrive at intersections. In intersections, PBLA attempts to find an optimal path to the destination by using probabilities that gain by learning automata algorithm.

At the end, performance of PBLAwas evaluated by the simulation and compared with the GPSR and GPCR. It is shown that, our scheme is able to increase reliability and decrease the number of hopsby using traffic information in learning phase. Moreover, contrary to topology-based, PBLA did not need to pre-determined route.

Synchronize the probability vector of all vehicles is one of the issues. We plan to investigate this issue in the future work. Another future work is scale protocol in big city map which city divided into different parts and protocol is applied on each part. In addition, learning in long term is not possible because of the simulator constraints. Whatever more increasing learning and getting street exceptions, PBLA could do more accurate routing.

**Authors**

FatemehTeymoori received the B.Sc. degree in Computer Software Engineering fromKashan Branch of Islamic Azad University, Kashan, Iran, in 2009. The M.Sc. degree inComputer Architecture Engineering from Department of Electrical and ComputerEngineering, ShahidBeheshti University, Tehran, Iran, in 2012. Her research interests include routing in vehicular ad hoc network.

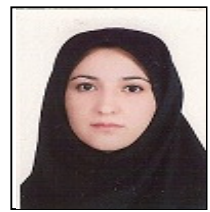

Hamid Nabizadeh received the B.Sc. degree in ComputerHardware Engineering fromShahidRajaee University, Tehran, Iran, in 2009. The M.Sc. degree inComputer Architecture Engineering from Department of Electrical and ComputerEngineering, ShahidBeheshti University, Tehran, Iran, in 2011. His research interests include security and routing in wireless sensor network and the data sharing in vehicular ad hoc network.

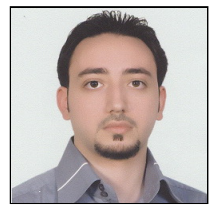

FarzanehTeymoori received the B.Sc. degree in Computer Software Engineering fromKashan Branch of Islamic Azad University, Kashan, Iran, in 2009. She is studying The M.Sc.inComputerArtificial Intelligent Engineering inScience and Research Branch of Islamic Azad University, Tehran, Iran. Her research interests include learning automata in the vehicular ad hoc network and in the image processing.

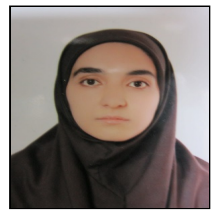